\documentclass[11pt]{article}
\usepackage{amsmath,amsfonts,amssymb}
\numberwithin{equation}{section}

\newcommand{\nc}{\newcommand}
\nc{\bib}{\bibitem}
\nc{\al}{\alpha}
\nc{\g}{\gamma}
\nc{\G}{\Gamma}
\nc{\D}{\Delta}
\nc{\eps}{\epsilon}
\nc{\vareps}{\varepsilon}
\nc{\la}{\lambda}
\nc{\La}{\Lambda}
\nc{\var}{\varphi}
\nc{\pa}{\partial}
\nc{\nn}{\nonumber \\ }
\nc{\be}{\begin{equation}}
\nc{\ee}{\end{equation}}
\nc{\bea}{\begin{eqnarray}}
\nc{\eea}{\end{eqnarray}}
\nc{\bra}[1]{\langle {#1}|}
\nc{\ket}[1]{|{#1}\rangle}
\nc{\sbar}{\bar{s}}
\nc{\Ab}{\bar{A}}
\nc{\Db}{\bar{D}}
\nc{\Lc}{\mathcal{L}}
\nc{\Oc}{\mathcal{O}}
\nc{\Qh}{\hat{Q}}
\nc{\ab}{\bar{a}}
\nc{\kappab}{\bar{\kappa}}
\nc{\Xib}{\bar{\Xi}}
\nc{\kb}{\bar{k}}
\nc{\rb}{\bar{r}}
\nc{\Rb}{\bar{R}}
\nc{\gh}{\hat{g}}
\nc{\Gh}{\hat{G}}
\nc{\kh}{\hat{k}}
\nc{\rh}{\hat{r}}
\nc{\sh}{\hat{s}}
\nc{\thh}{\hat{t}}
\nc{\phih}{\hat{\phi}}
\nc{\rhoh}{\hat{\rho}}
\nc{\Dh}{\hat{\Delta}}
\nc{\Kc}{\mathcal{K}}
\nc{\bb}{\mbox{\boldmath$b$}}
\nc{\db}{\mbox{\boldmath$d$}}
\nc{\fb}{\mbox{\boldmath$f$}}
\nc{\gb}{\mbox{\boldmath$g$}}
\nc{\hb}{\mbox{\boldmath$h$}}
\nc{\Kb}{\mbox{\boldmath$K$}}
\nc{\Qb}{\mbox{\boldmath$Q$}}
\nc{\Pb}{\mbox{\boldmath$P$}}
\nc{\Eb}{\mbox{\boldmath$E$}}
\nc{\Hb}{\mbox{\boldmath$H$}}
\nc{\Fb}{\mbox{\boldmath$F$}}
\nc{\Tb}{\mbox{\boldmath$T$}}
\nc{\Dbb}{\mbox{\boldmath$D$}}
\nc{\bbh}{\mbox{\boldmath$\hat{b}$}}
\nc{\fbh}{\mbox{\boldmath$\hat{f}$}}
\nc{\gbh}{\mbox{\boldmath$\hat{g}$}}
\nc{\hbh}{\mbox{\boldmath$\hat{h}$}}
\nc{\Kbh}{\mbox{\boldmath$\hat{K}$}}
\nc{\Qbh}{\mbox{\boldmath$\hat{Q}$}}
\nc{\pab}{\mbox{\boldmath$\pa$}}

\begin{document}

\topmargin -5mm
\oddsidemargin 5mm

\setcounter{page}{1}

\vspace{8mm}
\begin{center}
{\LARGE {\bf On hidden symmetries}}
\\[.3cm]
{\LARGE {\bf of extremal Kerr-NUT-AdS-dS black holes}}

\vspace{8mm}
 {\LARGE J{\o}rgen Rasmussen}
\\[.3cm]
 {\em Department of Mathematics and Statistics, University of Melbourne}\\
 {\em Parkville, Victoria 3010, Australia}
\\[.4cm]
 {\tt j.rasmussen@ms.unimelb.edu.au}

\end{center}

\vspace{8mm}
\centerline{{\bf{Abstract}}}
\vskip.4cm
\noindent
It is well known that the Kerr-NUT-AdS-dS black hole admits two linearly independent Killing 
vectors and possesses a hidden symmetry generated by a rank-2 Killing tensor.
The near-horizon geometry of an extremal Kerr-NUT-AdS-dS black hole admits four linearly 
independent Killing vectors, and we show how the hidden symmetry of the black hole itself is 
carried over by means of a modified Killing-Yano potential which is given explicitly.
We demonstrate that the corresponding Killing tensor of the near-horizon geometry
is reducible as it can be expressed in terms 
of the Casimir operators formed by the four Killing vectors. 
\renewcommand{\thefootnote}{\arabic{footnote}}
\setcounter{footnote}{0}

\newpage

\section{Introduction}

Isometries of spacetimes are generated by Killing vector fields, and one can associate a 
conserved quantity to each of these. A spacetime may also possess so-called hidden 
symmetries generated by higher-rank tensor fields. Following Carter's celebrated 
work \cite{Car68} on the Kerr black hole, it was realized \cite{WP70} that his new integral 
of motion is quadratic in momenta and generated by a rank-2 Killing tensor of the Kerr spacetime. 
Similar results have now been obtained for many classes of black holes and in various 
dimensions. We refer to the recent reviews \cite{ER0801,FK0802} for details and references.

Here we consider the family of extremal Kerr-NUT-AdS-dS black holes and our main focus is on the 
near-horizon geometries. Such geometries have attracted a lot of attention lately due to their role 
in the recently proposed Kerr/CFT correspondence \cite{GHSS0809}.
Extensive lists of references on this topic can be found in \cite{Ras1005}.

First, we review the Kerr-NUT-AdS-dS black hole in the metric form discussed in \cite{CLP0604}.
This spacetime admits two linearly independent Killing vectors and possesses a hidden 
symmetry generated by a rank-2 Killing tensor. It is also recalled \cite{KF07} that this Killing 
tensor follows from a Killing-Yano potential. We then determine the metric of the near-horizon 
geometry of the extremal Kerr-NUT-AdS-dS black hole
and verify that it satisfies Einstein's field equations with the same cosmological constant as for the
black hole itself. The isometry group, on the other hand, is enhanced as the 
near-horizon geometry admits four linearly independent Killing vectors. 
In the limiting procedure used to obtain the near-horizon geometry, the Killing-Yano 
potential diverges. However, a gauge freedom in the definition of the Killing-Yano potential 
allows us to construct a well-defined potential for the near-horizon geometry.
We finally demonstrate that the corresponding rank-2 Killing tensor is reducible as it can be 
expressed in terms of the Casimir operators formed by the four Killing vectors of the near-horizon
geometry.

\section{Kerr-NUT-AdS-dS black holes}

\subsection{Geometry}

Adopting the unit convention where $G=c=1$, a general Kerr-NUT-AdS-dS black hole of mass $M$ and with 
rotation parameter $a$ is described by the metric \cite{CLP0604}
\be
 d\sh^2=-\frac{\D_r}{\rh^2+y^2}\big(d\thh+y^2d\psi\big)^2+\frac{\D_y}{\rh^2+y^2}\big(d\thh-\rh^2d\psi\big)^2
  +\frac{\rh^2+y^2}{\D_r}d\rh^2+\frac{\rh^2+y^2}{\D_y}dy^2
\label{ds2alg}
\ee
where
\be
 \D_r=(\rh^2+a^2)\big(1+\frac{\rh^2}{\ell^2}\big)-2M\rh,\qquad
 \D_y=(a^2-y^2)\big(1-\frac{y^2}{\ell^2}\big)+2Ly
\label{Dy}
\ee
Here $L$ is the NUT parameter (see also \cite{GP0702}) and the determinant $\hat{g}$ of the metric 
only depends on $\rh$ and $y$ as
\be
 \sqrt{-\hat{g}}=\rh^2+y^2
\ee
This spacetime satisfies Einstein's field equations
\be
 \Gh_{\mu\nu}+\frac{3}{\ell^2}\gh_{\mu\nu}=0
\ee
and is AdS (dS) for positive (negative) renormalized cosmological constant $\ell^{-2}$. 
It reduces to Kerr for $L=\ell^{-2}=0$.
The horizons of the black hole are located at the positive zeros of $\D_r$, and the value of $\rh$ at the outer
horizon is denoted by $r_+$.
The isometry group $U(1)\times U(1)$ of (\ref{ds2alg}) is generated by the Killing vector fields
\be
 \{\pab_{\psi}\}\cup\{\pab_{\thh}\}
\ee

\subsection{Hidden symmetry}

As observed in \cite{KF07} (see also the reviews \cite{FK0802}), 
the metric (\ref{ds2alg}) admits the so-called Killing-Yano potential (1-form)
\be
 \bbh=\frac{y^2-\rh^2}{2}\db\thh-\frac{\rh^2y^2}{2}\db\psi
\label{bhat}
\ee
This implies the existence of the principal conformal Killing-Yano tensor \cite{Kas68}
\be
 \hbh=\db\bbh
\label{hbh}
\ee
and its Hodge dual, the Killing-Yano tensor \cite{Yano52}
\be
 \fbh=\ast\hbh=\ast\db\bbh
\ee
{}From this, one can construct a symmetric Killing tensor by contraction
\be
 \hat{K}_{\mu\nu}=\hat{f}_{\mu\lambda}\hat{f}_{\nu}^{\phantom{\nu}\lambda}
\label{Kh}
\ee
This tensor $\Kbh$ is responsible for the hidden symmetry associated with the conserved quantity 
$\hat{K}^{\mu\nu}\hat{p}_\mu\hat{p}_\nu$ quadratic in the momenta $\hat{p}_\mu$.
Considering the construction of $\Kbh$ above, it appears natural to refer to the entire quadruplet $(\bbh,\hbh,\fbh,\Kbh)$ 
as the generator of the hidden symmetry.

It is straightforward to verify that 
\be
 \Kbh=\Qbh+\rh^2\gbh
\ee
where $\gbh$ is the metric tensor while the components of $\Qbh$ in the ordered basis $\{\thh,\rh,\psi,y\}$ 
are given by
\be
 [\hat{Q}_{\mu\nu}]=\left(\!\!\!\begin{array}{cccc} \D_r&0&\D_r y^2&0\\ 0&\displaystyle{-\frac{(\rh^2+y^2)^2}{\D_r}}&0&0\\
   \D_r y^2&0&\D_r y^4&0\\  0&0&0&0   \end{array}  \!\!\right)
\ee 
With raised indices, the components are
\be
 [\hat{Q}^{\mu\nu}]=\frac{1}{\D_r}\left(\!\!\begin{array}{cccc} \rh^4&0&\rh^2&0\\ 0&-\D_r^2&0&0\\
   \rh^2&0&1&0\\  0&0&0&0   \end{array}  \!\!\right)
\ee

\section{Extremal Kerr-NUT-AdS-dS black holes}

When the inner and outer horizons of the black hole (\ref{ds2alg}) coalesce, the black hole is said to be extremal.
This happens when the otherwise single pole $r_+$ of $\D_r$ is a double pole in which case
\be
 \D_r(r_+)=\D'_r(r_+)=0
\label{DhDh}
\ee
We denote the value of $\rh$ at this single horizon by $\rb$.
The conditions (\ref{DhDh}) can be used to express $M$ and $a^2$ at extremality in terms of $\rb$ 
\be
 M=\frac{\rb\big(1+\frac{\rb^2}{\ell^2}\big)^2}{1-\frac{\rb^2}{\ell^2}},
   \qquad
 a^2=\frac{\rb^2\big(1+\frac{3\rb^2}{\ell^2}\big)}{1-\frac{\rb^2}{\ell^2}}
\ee

\subsection{Near-horizon geometry}

To describe the near-horizon geometry of an extremal Kerr-NUT-AdS-dS black hole,
we introduce (in the spirit of \cite{BH9905}) the new coordinates $t,r,\phi$
\be
 t=\frac{(\rb^2+a^2)\eps\thh}{\rb^2\rb_0},\qquad
 r=\frac{\rh-\rb}{\eps\rb_0},\qquad
 \phi=\rb^2\psi-\thh
\label{subs}
\ee
where
\be
 \rb_0^2=\frac{(\rb^2+a^2)\big(1-\frac{\rb^2}{\ell^2}\big)}{1+\frac{6\rb^2}{\ell^2}
  -\frac{3\rb^4}{\ell^4}}
\label{rrho}
\ee
The transformation of the radial coordinate facilitates zooming in on the near-horizon region while
the accompanying transformations ensure that the line element (\ref{ds2alg}) is well-defined in the
limit $\eps\to0$. The near-horizon geometry is then obtained by taking the limit $\eps\to0$ in which case
the metric becomes\footnote{Using different coordinates, a similar limit and near-horizon geometry is discussed in 
\cite{Ghe0902}, although Einstein's field equations (\ref{Gg}) are not addressed there.}
\be
 d\bar{s}^2=\frac{\rb_0^2(\rb^2+y^2)}{\rb^2+a^2}\big(-r^2dt^2+\frac{1}{r^2}dr^2\big)
  +\frac{\D_y}{\rb^2+y^2}\big(d\phi+\kb rdt\big)^2+\frac{\rb^2+y^2}{\D_y}dy^2
\label{ds2Near}
\ee
where
\be
 \kb=\frac{2\rb\rb_0^2}{\rb^2+a^2}
\label{kb}
\ee
It is straightforward, albeit tedious, to verify explicitly that this metric satisfies Einstein's field 
equations
\be
 G_{\mu\nu}+\frac{3}{\ell^2}g_{\mu\nu}=0
\label{Gg}
\ee
We find that the determinant $g$ of the metric (\ref{ds2Near}) only depends on the coordinate $y$ as
\be
 \sqrt{-g}=\frac{\rb_0^2(\rb^2+y^2)}{\rb^2+a^2}
\label{g}
\ee
and it is noted that the metric describes the near-horizon geometry of the Kerr black hole when $N=\ell^{-2}=0$.

The exact isometry group of the spacetime (\ref{ds2Near}) 
is generated by the Killing vectors
\be
 \big\{\Pb=\pab_\phi\big\}\cup\big\{\Tb=\pab_t,\ \Dbb=t\pab_t-r\pab_r,\ \Fb=\big(t^2+\frac{1}{r^2}\big)\pab_t
   -2tr\pab_r-\frac{2\kb}{r}\pab_\phi\big\}
\label{iso}
\ee
It follows that the isometry group $U(1)\times U(1)$ of the Kerr-NUT-AdS-dS black hole is enhanced to
$U(1)\times SL(2)$ in the near-horizon geometry.

\subsection{Hidden symmetry}

After the substitution (\ref{subs}), the Killing-Yano potential $\bbh$ (\ref{bhat}) {\em diverges} in 
the limit $\eps\to0$. This could indicate that the near-horizon geometry does not possess
a hidden symmetry admitting a 
Killing-Yano potential, at least not one carried over from the original Kerr-NUT-AdS-dS
spacetime. This is not the case, though, as the divergence problem can be resolved 
and a well-defined Killing-Yano potential can be constructed.
To demonstrate this, we use that a Killing-Yano potential is defined only up to a {\em constant} 
1-form. This corresponds to gauge invariance of the corresponding Killing-Yano and Killing 
tensors. 
We thus find that
\be
 \bbh+\tfrac{1}{2}\big(\alpha\db\thh+\beta\db\psi\big)
\ee
with $\alpha$ and $\beta$ constant, remains well-defined in the limit $\eps\to0$ provided
\be
 -\rb^4+\alpha \rb^2+\beta=0
\ee
The 1-form thereby obtained is given by
\be
 \bb=-\frac{\kb(\rb^2+y^2)r}{2}\db t-\frac{y^2}{2}\db\phi
\ee
and we have verified explicitly that the tensor $\Kb$ constructed as in (\ref{hbh})-(\ref{Kh}) by
\be
 \hb=\db\bb,\qquad
 \fb=\ast\hb=\ast\db\bb,\qquad
  K_{\mu\nu}=f_{\mu\lambda}f_{\nu}^{\phantom{\nu}\lambda}
\ee
is indeed a Killing tensor.
The 1-form $\bb$ is therefore the sought-after Killing-Yano potential for the Killing tensor $\Kb$.
We conclude that the quadruplet $(\bb,\hb,\fb,\Kb)$ generates a hidden symmetry of the near-horizon 
geometry (\ref{ds2Near}) of a Kerr-NUT-AdS-dS black hole.

It is observed that the Killing tensor $\Kb$ decomposes as
\be
 \Kb=\Qb+\rb^2\gb
\label{KQg}
\ee
where $\gb$ is the metric tensor of the near-horizon geometry (\ref{ds2Near}) while the components of $\Qb$ are
given by
\be
 [Q_{\mu\nu}]=-\sqrt{-g}\left(\!\!\begin{array}{cccc} 
  -r^2&0&0&0\\ 
  0&\displaystyle{\frac{1}{r^2}}&0&0\\
  0&0&0&0\\ 
  0&0&0&0   \end{array}  \!\!\right)
\ee 
Aside from the $y$-dependent factor $\sqrt{-g}$, this is recognized as the metric tensor of $AdS_2$ in the
coordinates $t,r$. With raised indices, the components of $\Qb$ only depend on the radial coordinate and are given by
\be
 [Q^{\mu\nu}]=\left(\!\!\begin{array}{cccc} 
  \displaystyle{\frac{\rb^2+a^2}{\rb_0^2 r^2}}&0&\displaystyle{-\frac{2\rb}{r}}&0\\ 
  0&\displaystyle{-\frac{(\rb^2+a^2)r^2}{\rb_0^2}}&0&0\\
  \displaystyle{-\frac{2\rb}{r}}&0&\displaystyle{\frac{4\rb^2\rb_0^2}{\rb^2+a^2}}&0\\ 
  0&0&0&0   \end{array}  \!\!\right)
\ee 
Since the metric tensor is multiplied by a {\em constant} in the decomposition (\ref{KQg}), the `essential' part of the hidden
symmetry is generated by $\Qb$ implying that the metric-tensor term can be ignored when
analyzing the hidden symmetry generated by $\Kb$.

We find that this hidden symmetry is {\em reducible} in the sense that $Q^{\mu\nu}$ can be expressed in terms of
the Killing vectors (\ref{iso}) by
\be
 Q^{\mu\nu}=\gamma
     \big(\Tb\otimes\Fb+\Fb\otimes\Tb-2\Dbb\otimes\Dbb\big)^{\mu\nu}
  -\frac{2}{\gamma\rb^2}(\Pb\otimes\Pb)^{\mu\nu}
\label{Qgamma}
\ee
where
\be
 \gamma=\frac{\rb^2(1+\rb^2/\ell^2)}{\rb_0^2(1-\rb^2/\ell^2)}
\ee
The Killing tensor $\Qb$ is thus a linear combination of the quadratic Casimir operators formed out of the two sets of
Killing vectors (\ref{iso}).
The various constants appearing in this decomposition can be absorbed by rescaling the Killing vectors.

\subsubsection*{Note added}
\vskip.1cm
\noindent
During the completion of this work, the paper \cite{Gal1009} appeared.
It has some overlap with the present work as it also discusses the hidden symmetry of the 
near-horizon geometry of the extremal Kerr black hole. The reducibility of the corresponding Killing tensor is
discussed using Poisson brackets in the context of particle dynamics near the extremal Kerr throat.
The paper \cite{Gal1009}
does not, however, consider the near-horizon geometry of the general Kerr-NUT-AdS-dS black hole, 
and it does not address the Killing-Yano potential underlying the hidden symmetry.
On the other hand, it proposes an ${\cal N}=2$ supersymmetric extension of the particle dynamics 
and it may be of interest to generalize this to the Kerr-NUT-AdS-dS scenario considered here.


\subsection*{Acknowledgments}
\vskip.1cm
\noindent
This work is supported by the Australian Research Council.


\end{document}